\documentstyle[12pt,epsf,psfig,rotate]{article}
\def\lsim{\mathrel{\rlap {\raise.5ex\hbox{$ < $}}
{\lower.5ex\hbox{$\sim$}}}}

\newcommand{\pr}{\paragraph{}}
\newcommand{\be}{\begin{equation}}
\newcommand{\ee}{\end{equation}}
\newcommand{\bea}{\begin{eqnarray}}
\newcommand{\nn}{\nonumber}
\newcommand{\eea}{\end{eqnarray}}

\newcommand{\nk}{\noindent}
\begin{document}
 
\begin{titlepage}
\begin{flushright}
OUTP--98--51P \\
hep-ph/9807451 \\
\end{flushright}
\begin{centering}
\vspace{.3in}
{\large {\bf High-Energy QCD as a Topological Field Theory  }} \\
\vspace{.3in}
{\bf John Ellis$^{a}$} and
{\bf N.E. Mavromatos$^{b}$}
\\
\vspace{.3in}
\vspace{.1in}
{\bf Abstract} \\
\vspace{.05in}
\end{centering}
{\small  We propose an identification of the conformal field theory
underlying
Lipatov's spin-chain model of high-energy scattering in perturbative
QCD. It is a twisted $N=2$ supersymmetric topological field theory, which
arises as the limiting case of the $SL(2,R)/U(1)$ non-linear
$\sigma$ model that also plays a role in describing the Quantum
Hall effect and black holes in string theory.
The doubly-infinite set of non-trivial integrals
of motion of the high-energy spin-chain model
displayed by Faddeev and Korchemsky are
identified as the Cartan subalgebra of a
$W_{\infty}  \otimes   W_{\infty}$
bosonic sub-symmetry possessed by this topological theory. The
renormalization group and an analysis of instanton perturbations
yield some understanding why this particular topological spin-chain
model emerges in the high-energy limit, and provide a new estimate
of the asymptotic behaviour of multi-Reggeized-gluon exchange.}

\begin{flushleft}
\end{flushleft}
\vspace{0.3in}
$^a$ Theory Division, CERN, CH 1211 Geneva 23, Switzerland,  \\
$^b$ P.P.A.R.C. Advanced Fellow, University of Oxford, 
Dept. of Physics (Theoretical Physics), 
1 Keble Road OX1 3NP, U.K. \\

\vspace{0.3in}
\begin{flushleft}
OUTP--98--51P \\
July 1998 \\
\end{flushleft}
\end{titlepage}
\newpage
\section{Introduction and Summary}
\pr
    In the last few years, Lipatov~\cite{Lipatov} and others
have developed a theory
of high-energy scattering in perturbative QCD, based on the t-channel
exchanges of Reggeized gluons interacting via s-channel gluons.
In the large-$N_c$ limit, the elastic scattering amplitude is related
to eigenstates of Hamiltonians with nearest-neighbour interactions,
that are holomorphic and antiholomorphic functions of the transverse
coordinates, and whose eigenvalues determine the asymptotic
behaviour in the high-energy limit. These Hamiltonians
for the exchange of $N_g$ Reggeized gluons
can be
written as :
\be
H_{N_g} = \sum _{k=1}^{N_g} H_{k,k+1} \qquad ; \qquad
{\overline H}_{N_g} =
\sum _{k=1}^{N_g} {\overline H}_{k,k+1}
\label{one}
\ee
One imposes  
periodic boundary conditions $H_{n,n+1}$ = $H_{n,1}$ in the
holomorphic sector, and analogously ${\overline H}_{n,n+1}$ =
${\overline H}_{n,1}$ in the
anti-holomorphic sector, where we use bars to denote the
replacements $z \rightarrow {\overline z}$, etc.. The 
two-particle interactions can be expressed
in several equivalent forms :
\bea
H_{jk} &=& P^{-1}_jlog(z_j - z_k)P_j + P^{-1}_k log(z_j - z_k)P_k
+ 2 \gamma _E \nn \\
~& = & 2log(z_j - z_k) + (z_j - z_k) log(P_jP_k) (z_j - z_k)^{-1}
+ 2\gamma _E \nn \\
~ & = & \sum_{l=0}^{\infty} (\frac{2l + 1}{l(l+1) - {\hat L}_{ik}^2}
- \frac{2}{l + 1})
\label{two}
\eea
where
\be
P_i \equiv i \frac{\partial}{\partial z_i} , \qquad
 {\hat L}_{ik}^2 \equiv (z_i - z_k)^2P_iP_k
\label{three}
\ee
and $\gamma_E$ is the Euler constant. Lipatov conjectured~\cite{Lipatov} 
that
the model was integrable, was able to solve
the case of two-gluon exchange exactly, suggested that the general
case could be solved using the Bethe Ansatz, and exhibited some
non-trivial integrals of motion.
\pr
Faddeev and Korchemsky~\cite{fadeev} have observed that the Lipatov
Hamiltonians are just those of Heisenberg ferromagnets with
non-compact spins $s$ = 0, -1.
This observation is prompted by the fact
that Lipatov's kernel ${\hat L}_{ij}$ can be represented as
a Heisenberg interaction term
\be
   {\hat L}_{ij} = S_i \cdot  S_j
\label{heisen}
\ee
among spin operators
$S_i$ at neighboring sites $i,j$ of a chain, whose components
are defined
in the (anti-)holomorphic sector of impact parameter space as
follows :
\be
   S_i^{+}=z_i^2 \partial _i - s z_i \qquad ; \qquad
S_i^{-}=-\partial _i \qquad ; \qquad
S_i^3 = z_i \partial _i - s
\label{spins}
\ee
This identification
enabled them to find a doubly-infinite set of non-trivial
integrals of motion, verify the integrability of the model, and
solve it in the $N_g$ = 2 case using a generalized Bethe 
Ansatz~\cite{fadeev}.
However, they did not give any symmetry origin for the integrals
of motion, and neither they nor Lipatov~\cite{Lipatov} has identified the
specific two-dimensional field theory corresponding to this
lattice model in the large-$N_g$ limit.
\pr
Although the prototypical case $N_g = 2$ can teach us many important
lessons, it cannot by itself control the high-energy behaviour of QCD,
in particular because it does not respect unitarity. It is therefore
of interest to extend
the above-mentioned analyses of the $N_g = 2$ case to larger $N_g$.
There has indeed been
considerable work on the kernel for $N_g = 3$, in connection
with the odderon in QCD~\cite{odderon}, and the general case of large
$N_g$ has also been examined~\cite{braun}. Transitions in the $t$
channel
between different numbers of gluons have also been analyzed~\cite{2to4}.
It
would be valuable to develop a more powerful approach to the
analysis of the large-$N_g$ limit, which should 
resemble a two-dimensional field theory.
As a step towards this goal, in this paper we identify the two-dimensional
conformal
field theory underlying Lipatov's two-body spin-chain
Hamiltonian in the continuum limit.
\pr
In Section 2, we use symmetry principles and the
field-theoretical description~\cite{affleck} of analogous 
compact Heisenberg ferromagnetic
spin chains as guides to this identification. It is well known that
such systems correspond to compact $O(3)$ non-linear $\sigma$ models
in the unitary condensed-matter cases ${s = 1/2, 1, 3/2, ...}$.
When the number of spin carriers at each site is fixed, such spin
chains possess a local $U(1)$ symmetry, and we demonstrate that this
is also a property of Lipatov's scattering kernel~\cite{Lipatov} for
Reggeized
gluons, and of his effective Hamiltonian. 
\pr
We argue in Section 3 that
the conformal field theory corresponding to high-energy QCD is the
limiting case $s \rightarrow 0^-$ (and $s \rightarrow -1^+$) of the
non-compact ${SO(2,1)/U(1)}$ non-linear
$\sigma$ model, which describes a Heisenberg system with ${s < 0}$
and is known to possess a $W_{\infty}$ symmetry. The doubly-infinite
set of conserved quantities exhibited by Faddeev and 
Korchemsky~\cite{fadeev} 
is
the Cartan subalgebra of a $W_{\infty}  \otimes  W_{\infty}$
symmetry that appears in the continuum limit. This is a bosonic
subalgebra of a twisted ${N = 2}$ supersymmetric $W$ algebra
possessed by the topological theories that are the
limits of
the non-compact non-linear $\sigma$ models when
${s \rightarrow 0, -1}$, which
correspond in the continuum limit
to high-energy scattering in
perturbative QCD. Quantum Hall conductors~\cite{Hall} and stringy black
holes~\cite{Witten}
are known to be described by the same non-compact non-linear
$\sigma$ model for generic values of the Landau level filling
parameter $\nu$ (black hole mass), and to possess this enhanced
${N = 2}$ supersymmetry in the limit of complete filling
${\nu = -1/s = 1}$ (at the core of the black hole).
\pr
As we show in Section 4, instantons play an important r\^ole in the
renormalization-group flow that drives the non-compact
$\sigma$ model towards the limiting case $s \rightarrow 1$.
They also provide us with a parametric estimate of the dependence
of the high-energy behaviour of the exchange of a large
number of Reggeized gluons, corresponding to a cylindrical
topology for the system exchanged in the $t$ channel. More
complicated topologies for the exchanged system could also be
treated within this field-theoretical approach, but are not discussed in
this paper.
\pr
Finally, in Section 5 we summarize our conclusions, emphasize the
aspects in which our analysis stands in need of confirmation,
and appraise some of the prospects for future progress.

\section{Symmetries of Lipatov's Spin-Chain Model}
\pr
It will be convenient for our subsequent discussion
to expand the Hamiltonian (\ref{two}) formally
as an infinite series in powers of the Heisenberg operator
(\ref{heisen}) $S_i \cdot S_j $,
\be
H_{ij} =
-\frac{1}{S_i \cdot S_j } + {\rm const} +
\sum _{l=0}^{\infty} \frac{2l + 1}
{l^2(l+1)^2} (S_i \cdot S_j ) + O[(S_i \cdot S_j )^2]
\label{expn}
\ee
where the omitted operators are higher powers of the Heisenberg
operator. They constitute irrelevant
operators in a renormalization group sense, by naive power counting,
so do not affect the continuum limit of the theory, and we shall
not deal explicitly with them in what follows.
However (\ref{expn})
does contain a non-analytic term ${1/S_i \cdot S_j}$,
due to the $l=0$ partial wave
in (\ref{two}), for which the Taylor expansion fails.
This is a particular
feature of the non-compact spin formalism,
and should be contrasted to the conventional  case of
compact spin $s > 0$,
where Heisenberg chains can be
represented as non-singular functions
of $S_i \cdot S_j$.
Formally, as we shall describe below,
one can regularize the $l=0$ limit
by representing the finite-size spin chain, which corresponds
to a fixed number of gluons $N_g$, as an infinite-size 
lattice chain with `holes', i.e., missing spins.
As a quantum-mechanical problem, removing spins is a complicated
procedure since it involves modification of the Hilbert space.
However,
attempts
have been made
to describe doping in
anti-ferromagnetic chains,
with the hope of understanding
the relevance of the hole dynamics in
scenarios for magnetic superconductivity~\cite{shankar,dorey}.
Below we borrow from these techniques.
\pr
An important feature of
the model (\ref{one}, \ref{two}) is the fact that
the number of Reggeized gluons per lattice site
is fixed. This implies that
there is a local gauge symmetry in the Heisenberg
interaction $S_i \cdot S_j$, which simply expresses
the particle-number conservation law.
This symmetry can be seen straightforwardly
if we represent the Heisenberg interaction
in terms of fundamental `Reggeon' creation and annihilation
operators
$C^\dagger _{\alpha,i}$ and
$C_{\alpha,i}$,
in direct analogy with the corresponding representation
in the solid-state models relevant to the description of
high-temperature
superconductivity~\cite{dorey}. To this end, we write the
Heisenberg spin-spin interaction in a `microscopic' form
\be
     S_i \cdot S_j = -J \sum_{\langle ij \rangle} \sum_{\alpha,\beta}
T_{i,\alpha\beta} T_j^{\alpha\beta} \qquad ; \qquad
T^{\alpha\beta}_i \equiv C^{\dagger, \alpha}_i C^\beta
_i~({\rm no~sum~over~i})
\label{tab}
\ee
where $\alpha,\beta$ denote spin $s$ indices, $i,j$ are lattice-site
indices
and $\langle \dots \rangle$ denote nearest-neighbour sites.
To exhibit the $U(1)$ symmetry we introduce a
{\it slave-boson} Ansatz
\be
C_i^\alpha = \psi _{i,\alpha} z^\dagger
\label{sfa}
\ee
where $\psi _{\alpha,i}$, $\psi _{\alpha,i}^\dagger $
are fermion operators that annihilate or
create a `hole'
at a site $i$ in the spin representation $s$. They
satisfy
canonical commutation relations. On the other hand, the $z$,
$z^\dagger$
are
Bose fields that are spin singlets. The Ansatz (\ref{sfa})
satisfies trivially a local
$U(1)$
phase
symmetry
\bea
\psi _{j,\alpha}
\rightarrow e^{i\theta_j} \psi _{j,\alpha} \nn \\
 z _j
\rightarrow e^{i\theta_j} z _j
\label{gauge}
\eea
which is a consequence of the local
constraint restricting the number of Reggeons per site :
\be
C_i^\dagger C_i =\psi^\dagger _i \psi_i + z^\dagger z
= 2s
\label{number}
\ee
It can be shown~\cite{shankar,dorey} that
the above formalism is a convenient way of representing
the effects of holes in a spin chain as a path integral
over fermionic variables parametrizing the
Berry-phase term that describes the missing-spin effect in the
action. Essentially, one describes the effects
of the missing spin by subtracting from the action the contribution
that a spin would make if it were there. This is a simple
way of
describing
correctly
the change in the Hilbert space.
\pr
By implementing the slave-fermion Ansatz, one can
consider a situation where the total
number of Reggeons is fixed at
$N_g$, but the size of the lattice chain is infinite.
This fixed-Reggeon-number case
has a natural doping interpretation
and the infinite-chain limit provides
a field-theory interpretation in the continuum limit, as
we discuss in more detail in Section 3.
The `doping concentration' can be defined
formally as the vacuum expectation value $\eta$
of the fermion bilinears in a splitting
of the form :
\be
\psi ^\dagger_{\alpha,i} \psi _{\alpha,i} = <
\psi ^\dagger_{\alpha,i} \psi _{\alpha,i }> + :
\psi ^\dagger_{\alpha,i} \psi _{\alpha,i}     :
\equiv \eta \; +
: \psi ^\dagger_{\alpha,i} \psi _{\alpha,i}     :
\label{doping}
\ee
The splitting (\ref{doping}) provides us with the
advertized regularization of the non-analytic terms
$1/{S_i \cdot S_j}$ in (\ref{expn}). Using the constraint
(\ref{number}) and the appropriate free-fermion commutation relation,
the Heisenberg terms can be written in the form
\be
  S_i \cdot S_j  = \eta \; + : \psi ^\dagger _{\alpha, i}\psi _{\alpha, i}:
-  4s^2
\psi _{\alpha,j}\psi _{\alpha,i}^\dagger
\psi _{\beta,i}\psi _{\beta,j}^\dagger + \dots
\label{4fermi}
\ee
where the $\dots$ indicate terms that are irrelevant
operators in a renormalization group sense, by naive power counting.
One can expand (\ref{4fermi}) formally
in a Taylor series
in powers of ${4s^2}/{\eta}$,
and at the end
one can take the twin limit $\eta, s \rightarrow 0$
to recover the `half-filled' action (\ref{two}).
Clearly, the crucial test of the validity of this
`hole-regulator' scheme will be provided
by a careful study of the scaling properties of the model,
as a function of the doping concentration $\eta$. This is
left for future work. What we argue below is that
the above scheme provides one with the necessary tools
to study the symmetries of the model in a straightforward
and physical way.
\pr
Concentrating on
the relevant operators (by naive power counting)
one observes that
the gauge-invariant effective Hamiltonian
thus constructed contains amongst others a term
\be
   \sum_{ij}  J_{ij} \psi _{i,\alpha}^\dagger \psi _{j}^\alpha
\Delta _{ij} + \dots
\label{eff}
\ee
where $J_{ij}=({4s^2}/{\eta}) + \sum _{l=0}^{\infty}
{2l + 1}/[l^2(l+1)^2]$  and the link variable
$\Delta _{ij} =<\psi _i^{\alpha,\dagger}\psi _{\alpha,j}>$
is a Hubbard-Stratonovich gauge field.
For details of this construction
we refer the interested reader to the relevant
literature~\cite{dorey}.
Thus, the Hamiltonian can be written
in terms of the field
\be
    M_{ij}^{\alpha\beta} \equiv \psi _i^{\alpha\dagger} \psi _j^\beta
\label{string}
\ee
which is manifestly gauge invariant in the
light-cone gauge, for which $\Delta _{ij} =1 $ along the links.
This is a particularity of
a two-dimensional gauge theory, which
will provide us with the generators of an
infinite-dimensional Lie algebra of symmetries,
as we shall see later.
\pr
The basic advantage of the $U(1)$ gauge symmetry is that
it allows for a parafermion construction of the conformal
field theory corresponding to the continuum limit of the
statistical model. This is important because
the non-compact spin case of Lipatov's kernel
apparently corresponds to the limiting case of
a non-compact $SL(2,R)$
Wess-Zumino
$\sigma$ model, as we discuss in Section 3.
The extra $U(1)$ symmetry at any doping concentration,
i.e., arbitrary but fixed Reggeon number with at most 
one Reggeon present per site,
implies that one can mod out the local
Abelian phase factors to obtain an
$SL(2,R)/U(1)$ coset model. Such
models are equivalent to parafermion models,
and are known~\cite{bakir} to possess an infinite-dimensional
$W_\infty$
algebra of symmetries.
Below we construct such symmetries explicitly in the
lattice model.
\pr
Before doing so, we remark that Heisenberg chains with holes
are known~\cite{susy} to possess graded (supersymmetric) algebras
generated by particles (spins) and holes (superpartners).
To introduce the empty sites, one allows for a hopping element
\be
t_{ij} \psi _i ^\dagger \psi _j
\label{hopping}
\ee
in the Hamiltonian of the chain\footnote{In the 
gauge-invariant formalism discussed above, this hopping element
is induced by the doping and is accompanied by
the replacement $t_{ij} \rightarrow \Delta _{ij} t_{ij}$.}.
At half-filling, $t_{ij} \rightarrow
0$, and
this limit can be taken at the very end of our computation
if necessary, after obtaining important symmetry information,
e.g. on supersymmetry .
In terms of projection operators $\chi ^{AB} \equiv |A> <B| $
on the $2s +1$ states on a lattice site, including the empty ones,
the Hamiltonian reads
\be
H=\sum _{\sigma} \sum _{ij} (t_{ij}\chi _i^{0\sigma}\chi_j^{\sigma0}
 + J_{ij}
 \chi _i^{\sigma\sigma '}
 \chi _j^{\sigma\sigma '} )
\label{hamilt}
\ee
where the indices $\sigma$ denote spin states, with the exclusion of the empty ones.
\pr
The operators  $\chi ^{AB}$ satisfy a supersymmetry
algebra in the spin $s$ representation. In the non-compact
$s=-1$ case of interest such a supersymmetry becomes twisted,
and the holes represent ghost states.
Supersymmetry implies a $W_\infty \otimes W_{\infty} $
algebra in the bosonic sector~\cite{WW}.
A similar $W_\infty
\otimes W_\infty$
structure arises in the quantum Hall system:
one $W_\infty$ appears at each Landau level, and is
associated with magnetic translation operators, whilst the other
mixes the various levels, and is associated with operators
that appear in the Hamiltonian of the model~\cite{Hall}. In the
Quantum Hall
case
there is also
an associated supersymmetry, similar to that of the
Heisenberg chain, which explains the underlying
$W_\infty \otimes W _\infty $
structure. The similarity of the Hall system
to the present model is discussed further
in the next Section, where it is argued
that both models can be mapped onto Wess-Zumino models
that belong to similar equivalence classes.
\pr
To see one of the $W_{\infty}$ structures, we
make use of the field (\ref{string}), in terms of which
Lipatov's kernel is expressed. One easily sees that
the following algebra is satisfied for the non-compact case
$s=-1$, which has a heighest-wight representation and
$|2s + 1|=1$ states :
\be
  [M_{i_1j_1} , M_{i_2j_2} ] =  \delta _{j_1 i_2}M_{i_1j_2}-
\delta_{j_2i_1}M_{i_2j_1}
\label{w2}
\ee
which is a field-theory realization of a
$W_\infty$ algebra. Notice the formal similarity
of the algebra (\ref{w2}) to the corresponding one generated
by fermion bilinears in two-dimensional large-$N_c$
QCD with adjoint fermions,
considered
in~\cite{mandal}. The difference
is that
in our case
Lipatov's Hamiltonian pertains to the pure `glue' sector
of quark-quark high-energy
scattering processes, and the associated fermion bilinears
arise from the mapping of the model to a spin system with doping.
\pr
The supersymmetry of the doped theory suggests
the existence of another $W_\infty$ structure.
This must be associated with the
bosonic degrees of freedom of the Ansatz (\ref{sfa}).
One can construct infinite bosonic symmetries out of these variables,
which resemble $W_\infty$ structures.
Indeed, the Hamiltonian (\ref{expn}) depends on the composite
bosonic bilocal operator $z_i z^\dagger _j $, which transforms like
a gauge link variable. In the continuum limit, one can define
the bilocal field  (in space)
\be
    \Phi (x,y; t) = z(x, t) z^\dagger (y, t)
\label{bos}
\ee
which satisfies
the $W_\infty$ algebra
\be
 [\Phi (x,y;t), \Phi (x',y'; t)] = \delta (x-y')\Phi (x,y')
- \delta (y' - x)\Phi (y,x')
\label{bosw}
\ee
it should be remarked that the above algebra is {\it classical}
in the sense that it was derived by operators in the model
which are constructed so as to obey the
canonical commutation relations. Quantum corrections
that arise after path integration \cite{dorey} should in general
modify the algebra by appropriate central  extensions, as
well as non-linear terms~\cite{bakir}.
Investigations of these issues falls beyond the scope
of the present article.
\pr
Notice that this algebra pertains to the
field space  of the two-dimensional high-energy limit of QCD.
In this respect it is a `target-space' symmetry algebra
of the corresponding $\sigma$ model.
It is clear from~\cite{fadeev}
that there is a double set of mutually-commuting
infinite-dimensional Cartan
subalgebras in Lipatov's model~\cite{Lipatov}. However, in contrast
to the above discussion, these
symmetries associated with the integrability
of the model are `world-sheet' symmetries.
The world sheet in this case
consists of the physical transverse impact-parameter space
of the Reggeized gluons, and its finite size is related to the number
of them in a physical high-energy quark-quark scattering process.
We expect that the connection between the target-space and
world-sheet pictures in this case
is provided in an analogous way to the elevation
of world-sheet $W_\infty$ symmetries to target space
in the two-dimensional black hole case. 
In the black-hole case, this association
was achieved by appropriate $(1,1)$ deformations, but
it remains to be seen what is the precise form of
such operators in our present case. Until this is done,
this form of association can only be considered as
a conjecture.
\pr
Before closing this Section, it is useful to
investigate the form of the
world-sheet symmetry algebras of the present model. Their
Cartan subalgebras
have been
constructed by
Fadeev and Korchemsky using Lax operator techniques~\cite{fadeev}.
Let us first concentrate in the $Q_k$ set, which in their
notation consists of operators of the form :
\be
Q_k = \sum_{n \ge i_1 \ge \dots i_k } i^k  z_{i_1i_2} z_{i_2i_3}
\dots z_{i_ki_1} \partial _{i_1}\dots \partial _{i_k}
\label{qk}
\ee
where $z_{i_1i_2} \equiv z_{i_1} - z_{i_2} $, $\partial _{i_1} \equiv
\partial /\partial z_{i_1} $.
A classical $w_{\infty}$ algebra acting on a holomorphic function
is generated by operators of the form
\be
   w_n^m \equiv z^m\partial _n
\label{wgen}
\ee
satisfying
\be
  [w_n^m, w_{n'}^{m'} ] = (nn' - m'm)w_{n + n'}^{m + m'} + \dots
\label{walg}
\ee
where the $\dots$ indicate possible quantum central extensions.
Viewing the commutator as a
 Poisson bracket on a two-dimensional phase space,
these are transformations that preserve the phase-space
area.
The Cartan subalgebra corresponds to the subset with
$m =n$, i.e., to an equal number
of coordinates and momenta, exactly as happens in (\ref{qk}).
One can, therefore, proceed formally
to construct the
remaining generators of the $w_\infty$ algebra by
defining
\be
Q_k^l = \sum_{n \ge i_l \ge \dots i_1 } i^k  z_{i_1i_2} z_{i_2i_3}
\dots z_{i_ki_1} \partial _{i_1}\dots \partial _{i_l}
\label{qkl}
\ee
These operators can be constructed explicitly
in the $s=0$ case,
and then extended to $s=-1$ by a similarity transformation.
The required transformation in our case is given by
\be
O_{s=-1} \equiv
(z_{12}z_{23}\dots z_{n1})^{-1}
O_{s=0}
(z_{12}z_{23}\dots z_{n1})
\label{simil}
\ee
These transformations apply in the case where there
are periodic boundary conditions in the finite-size chain.
For an infinitely-long doped chain the similarity
transformation is provided by
\be
S_{doped}^{-1} = \Pi _{i < k} (z_i - z_k)
\label{product}
\ee
Such similarity transformations are non-unitary and they also
appear in Quantum Hall systems, connecting operators
pertaining to the Integer and Fractional Quantum Hall Effects~\cite{Hall}.
The $W$ algebra pertaining to the operators (\ref{qkl})
corresponds to the algebra of a single Landau level in the
Quantum Hall case. We expect that
the second set, $I$, should correspond to the
$W$ algebra that mixes the Landau levels in the
Quantum Hall case, but this remains to be demonstrated.

\section{Generalized Heisenberg Ferromagnets, Compact and
Non-Compact Non-Linear $\sigma$ Models}
\pr
The precise nature of these world-sheet
$W$ algebras
can be investigated if one finds a $\sigma$-model
representation of the above picture, and uses
conformal field theory techniques to construct
the generators of the transformations.
As we shall argue below, the singular limit $s=0$
can be considered from the point of view of a non-compact spin problem,
which suggests the representation of the
theory in terms of a $SL(2,C)$ algebra.
Restricting ourselves to the subgroup $SL(2,R)$
and taking into account the extra $U(1)$ symmetry,
one can conjecture the form of the $\sigma$ model
that is appropriate for such a system:
it is a gauged Wess-Zumino model, based on the group
$SL(2,R)/U(1)$. The topological nature of the problem
may be captured by an appropriate twisted world-sheet
supersymmetry which can be taken to be $N=2$.
To substantiate these claims, but not to prove them rigorously,
we now review briefly the situation in
the compact spin case, and then 
continue the results analytically
to the non-compact case.
\pr
It is well known~\cite{affleck} that Heisenberg
spin
models may be mapped onto $O(3)$ non-linear ${\sigma}$-models
in the limit of large spin $s$,
\be
     S_1 ^2 + S_2 ^2 + S_3 ^2 = s ( s + 1 ) > 0
\label{aff0}
\ee
As reviewed in \cite{affleck},
the spin Hamiltonian for large spin $s$ corresponds to the
Lagrangian
\be
L= \frac{1}{2g} \partial _\mu \phi \partial ^\mu \phi
+ \frac{\theta}{8\pi} \epsilon ^{\mu\nu} \phi (\partial _\mu
\times \partial _\nu \phi )
\label{aff1}
\ee
with the conventional normalization
constraint $ {\bf \phi}^2 \equiv \phi _1^2
+ \phi _2^2 + \phi _3^2  $ = 1 and the following
identifications of the coupling constant $g$ and the topological
angle $\theta$ that appears in the antiferromagnetic case:
\be
        g=\frac{2}{s} \qquad, \qquad \theta = 2\pi s
\label{aff2}
\ee
We do not discuss further the interesting physics associated
with the $\theta$ parameter~\cite{affleck}.
In addition to the formal derivation of the $O(3)$
non-linear $\sigma$ model in the limit of large $s$, there is also
evidence that this model describes correctly features of
Heisenberg models with small $s$,
as reflected in the Figure. 
For example, both analytical
and numerical studies support the maintenance of N\'eel order in
the antiferromagnetic ground state for ${s = 1/2}$~\cite{affleck}, as assumed
in deriving the  $\sigma$ model in the continuum limit.

\begin{figure}[htb]
\begin{center}
\parbox[c]{8in}
{\psfig{figure=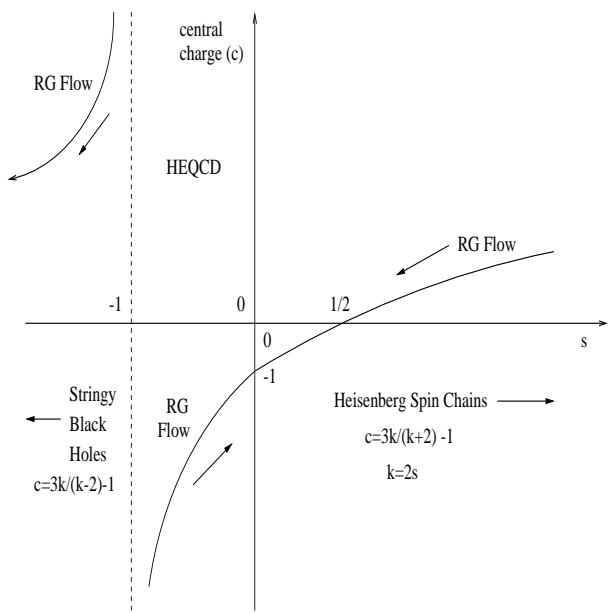,height=8cm,width=8cm}}
\end{center}
\caption{{\it Map 
of the space of the gauged compact and non-compact
non-linear $\sigma$ models discussed in this paper, including conventional
Heisenberg
spin chains with ${s \ge 1/2}$ that are described by $SO(3)/U(1)$ $CP^1$
models, and analytic continuations to ${s <0}$ that are described
by $SO(2,1)/O(2)$ or $SO(2,1)/O(1,1)$ models. High-energy
scattering in QCD is described by 
a twisted supersymmetric version of 
the limiting cases ${s = 0, -1}$,
and stringy black holes by models with ${s < -1}$.
The vertical coordinate measures the central charge $c$,
and we exhibit the renormalization-group 
flows towards ${s = 1/2}$ in the infrared limit of unitary
spin models, towards ${s = 0}$ for the anti-unitary models
whose twisted supersymmetric version describes 
high-energy scattering in perturbative QCD, and
away from ${s = -1}$ for the unitary stringy black-hole models.}}
\label{pdiag}
\end{figure}

\pr
The two-dimensional $O(3)$ non-linear $\sigma$ model is well
understood: indeed, its exact $S$-matrix is known. For our
purposes, the most useful formulation is as a
$CP^1$ $\sigma$ model, in which the unit vector ${\bf \phi}$
is represented using a two-component complex spinor $z_{\alpha}$:
\be
 \phi = z^{*} {\bf \sigma} z  \qquad ; \qquad |z|^2 = 1
\label{aff3}
\ee
where the $\sigma$ are Pauli matrices, with 
in terms of which the Lagrangian (\ref{aff1}) becomes
\be
L = \frac{1}{g} [ |\partial _\mu z |^2 +
(z^{*} \partial _\mu z )^2 ] + \frac{\theta}{2\pi}
\partial _\mu \epsilon ^{\mu\nu}(z^{*} \partial _\nu z)
\label{aff4}
\ee
Insight into this model is obtained by rewriting it in terms
of the $U(1)$ gauge field
\be
 A_\mu = i z^{*} \partial _\mu z
\label{aff5}
\ee
using which the Lagrangian (\ref{aff4}) can be expressed as
\be
L = \frac{1}{g} | (\partial _\mu + i A_\mu)z |^2 -
\frac{i\theta }{2\pi}\partial _\mu \epsilon ^{\mu\nu}A_\nu
\label{aff6}
\ee
Solving this model via a saddle-point method, one finds that
the $z$ bosons acquire masses
\be
m = \Lambda e^{-2 \pi /g} = \Lambda e^{- \pi s}
\ee
where $\Lambda$ is an ultra-violet cutoff. The $z$ bosons
are free, apart from interactions via a massless Abelian
gauge field with an effective gauge coupling strength
\be
e^2 = 6 {\pi}^2 m^2
\ee
This gauge interaction confines the $z$ bosons in the
${1+1}$-dimensional case, leading to a massive triplet of
bound states, in agreement with the exact $S$-matrix results
for the $O(3)$ non-linear $\sigma$ model~\cite{affleck}.
\pr
To identify the conformal field theory to which this
model corresponds one has to apply finite-size
scaling methods in the way studied in ref. \cite{affleck}.
The result of such an analysis indicates that
the $CP^1$ model belongs to the same equivalence
class as the $SU(2)$ Wess-Zumino conformal field theory.
As is well known~\cite{affleck}, the central charge of the $SU(2)$
Wess-Zumino model
is
\be
c = { 3 k \over (k + 2) }
\label{twelve}
\ee
where the level parameter ${k = 2 s}$. Note, in addition, that
$c$ is reduced by $1$ if a $U(1)$
subgroup is gauged.
One frequently considers an ultra-violet limit $\Lambda$
$\rightarrow$ $\infty$ in which the non-linear $\sigma$-model
coupling ${g = 2/s}$ also $\rightarrow$ $\infty$ (and hence $k \rightarrow 0$)
in such a way that the $z$-boson mass
$m$ and the $CP^1$
$U(1)$ gauge coupling $e$ remain fixed. However,
one can also consider the case in which $s = k/2$ is fixed
to be zero, corresponding formally to $g$ = $\infty$, and
take the limit $\Lambda$ $\rightarrow$ $\infty$. In this case,
$m$ and $e$ become infinite and the only remaining physical
field is the $U(1)$ gauge field, which is however completely
topological in nature, since it is non-propagating.
This interpretation of the ${s = k/2 = 0 }$
theory is supported by the fact that the formula (\ref{twelve})
yields a
central charge ${c = 0}$ in this case,
corresponding to a
topological gauge theory. We therefore identify
the ${s=0}$ Heisenberg model
formally as a topological $U(1)$ gauge theory in
the continuum limit.
\pr
We now consider the continuation of the above results to ${s < 0}$.
It is clear that the quantum Heisenberg model cannot be
represented in terms of Hermitean operators when ${-1 < s < 0}$,
since the quadratic Casimir
\be
{S_1}^2 + {S_2}^2 + {S_3}^2 = s(s+1) < 0
\label{thirteen}
\ee
for this range of $s$. One must continue one or more of the
spin components $S_{1,2,3}$ to complex values, and the simplest
two inequivalent
possibilities are to take one or two of the components to be
anti-Hermitean. Taking a naive continuum limit in which each
of the spin components is replaced by a
conventionally-normalized
local field variable, the two inequivalent possibilities
are
\be
- {\phi_1}^2 - {\phi_2}^2 + {\phi_3}^2 = 1
\ee
and
\be
{\phi_1}^2 - {\phi_2}^2 + {\phi_3}^2 = 1
\ee
where each of the field components ${\phi_{1,2,3}}$ is
understood to be real. These both represent $SO(2,1)$ group
manifolds, but with different gaugings. Implementing the
manifold constraint by taking $\phi_3$ as the dependent
variable, one finds
\be
\phi_3 = \sqrt(1 + {\phi_1}^2 + {\phi_2}^2)
\label{case1}
\ee
and
\be
\phi_3 = \sqrt(1 - {\phi_1}^2 + {\phi_2}^2)
\label{case2}
\ee
in the two cases. If these models are gauged the
corresponding
gaugings are with respect to compact $O(2)$ and non-compact
$O(1,1)$ respectively. We have no formal proofs, but expect
that non-linear $\sigma$ models on the non-compact manifolds
$SO(2,1)/O(2)$ and $SO(2,1)/O(1,1)$ are
the continuum field theories corresponding to possible
continuations of the spin systems (\ref{thirteen})
to the range ${-1 < s < 0}$.
The central charges for these models are known to be
\be
c = { 3 k \over (k - 2)}  - 1
\label{eighteen}
\ee
where the level parameter ${k = - 2 s}$ for ${s < 0}$.
The subtraction of unity in (\ref{eighteen}) reflects 
the gauging of the non-linear $\sigma$ model.
\pr
The variant corresponding to the high-energy scattering problem
should be the non-linear $SO(2,1)/O(2)$ or $SU(1,1)/U(1)$ model.
This has the correct local symmetry corresponding to the
conservation of the number of Reggeons exchanged in the
$t$ channel, i.e., the number of spin variables per lattice
site in the Heisenberg spin chain. Moreover, it is known to
possess a $W_{\infty}$ symmetry for generic values of the level
parameter ${k = -2 s}$~\cite{bakir}. As pointed out by Faddeev and
Korchemsky~\cite{fadeev},
Lipatov's model~\cite{Lipatov} of high-energy scattering can be regarded
as a
combination of a holomorphic ${s = - 1}$, i.e., ${k = 2}$, spin
chain and an anti-holomorphic ${s = 0}$, i.e., ${k = 0}$, spin
chain, and these are related by a similarity transformation. We
have already argued that the ${s = 0 }$ model is a topological
$U(1)$ gauge theory, and shall now argue the same for the
${s = - 1}$ model, based on a previous analysis~\cite{bakir} of the
$SU(1,1)/U(1)$ model for ${k > 2}$, i.e., ${s < - 1}$.
\pr
The $k > 2$ $SU(1,1)/U(1)$ model
is known~\cite{Witten} to describe a black hole in string theory, with
mass proportional to ${1/{\sqrt(k - 2)}}$,
as shown in the Figure. As can be seen from
equation (\ref{eighteen}),
the central charge ${c > 2}$ in this range of $k$,
and the string black hole becomes critical when ${k = 9/4}$,
since then ${c = 26}$ in the absence of other degrees of freedom.
It has been pointed out~\cite{emntop} that this model is
described in the neighbourhood of the singularity
at the centre of the black hole by a topological $U(1)$ gauge
theory coupled to matter fields $(a,b)$ that parametrize the
appearance of space-time coordinates away from the singularity $w$:
\be
\int d^2z \frac{k}{4\pi} ( D_za.D_{\bar z}b + w\epsilon^{ij} F_{ij})
\label{wza}
\ee
In the limit ${k \rightarrow 2}$, all of space-time is absorbed
by the singularity, ${a,b \rightarrow 0}$, and the theory
becomes a pure topological $U(1)$ gauge theory without matter
fields. It has further been pointed out that this theory has ${N = 2}$
supersymmetry, and that the bosonic part of this symmetry algebra
includes a $W_{\infty}$ $\otimes$ $W_{\infty}$ algebra~\cite{emntop}.
\pr
Such a topological theory has been constructed in~\cite{emntop}
by twisting
the $N=2$ supersymmetric Wess-Zumino model on
$SL(2,R)/U(1)$, adding to its stress tensor a derivative of the
$U(1)$ current so as to ensure $c=0$ \cite{eguchi}.
It is known that
$N=2$ superconformal
theories may be constructed by adding and subtracting
a free boson~\cite{lykken}.
Unitarity is in general required in those constructions,
although it may be relaxed in some cases.
The central charge of this
supersymmetric coset construction is given by
\be
c = \frac{3k}{k - 2}
\label{supersym}
\ee
yielding 
the result $c=0$ as $k=2s \rightarrow 0^{-}$, which coincides 
with the
ungauged case. Thus, we now see that
the case $s \rightarrow 0$ is free from ambiguities, in the sense that
in both the limits $ s \rightarrow 0^{\pm}$ the central charge $c
\rightarrow 0$ in an unambiguous way. 
\pr
However, the limit $s+1 \rightarrow 0$ is ambiguous, since
the central charge of the $s \rightarrow -1^-$ 
black-hole theory has the limiting behaviour
$c \rightarrow + \infty$, whilst the central charge 
in the case $s \rightarrow -1^- $ has the
limiting behaviour $ c \rightarrow -\infty$. 
One would like to find some `principal value'
prescription to resolve this ambiguity.
Crudely speaking, this should be an `average' between the
limits $s \rightarrow 1^{\pm}$ that is
a conformal field theory with $c=0$.
Our guiding principle in formulating this
is the Bethe-Ansatz approach 
of~\cite{fadeev}, where it is observed that the $s=0$ case is formally
isomorphic to the appropriate version of the $s=-1$ case.  
This prescription can be achieved formally
by representing the $s=-1$ case (and the equivalent $s=0^-$ model) 
as a topological 
$\sigma$-model field theory on the world sheet, 
that corresponds to the impact-parameter space in
the present case of high-energy QCD.
The requisite topological $\sigma$ model may be constructed by
the appropriate twisting~\cite{eguchi} of an
$N=2$ supersymmetric $\sigma$ model, causing the
effective central charge of the theory to vanish
in the ungauged case~\footnote{Upon this twisting,
the supersymmetric partner fermion fields become BRST ghost fields.}.
As we shall see, an important aspect of this construction is
the emergence of a cigar-like metric, whose singularity 
describes the limiting case $s(s +1)=0$. 
A nice feature of this type of metric is that it is the 
limiting non-compact target-space-time case that admits instanton
solutions, as we discuss in the next section.
\pr
To justify this scenario
formally, we pursue in more detail
our spin-charge-separation formalism for the description of the 
antiferromagnet, according to which
the magnon sector $z$ is described
by the $CP^1$ continuum field theory.
In the absence of fermions, the latter would be equivalent to the $O(3)$ 
$\sigma$ model written in terms of the $\eta^\alpha $, $\alpha =1,2,3$
variables: $\eta = {\overline z} \sigma^\alpha z$, where the
$\sigma ^\alpha$ are the $2 \times 2$ Pauli matrices, with 
${\overline z} z={\rm const.}$ corresponding to the Casimir condition
$\sum _{\alpha =1}^{3} \eta ^{2,\alpha} = s(s+1)$. 
As was discussed in~\cite{dorey}, there is a
corresponding approximate formalism which
is valid in the presence of fermions, in a variant of the 
Heisenberg chain with next-to-nearest-neighbor interactions.
In this case, the Casimir constraint on the magnon fields $z$ reads:
\be
    {\overline z} z + \frac{1}{G} \psi ^\dagger \psi = {\rm const.} 
\label{constr} 
\ee
where $G \sim t'/J' \rightarrow \infty$ in the model of~\cite{dorey},
where the primes denote next-to-nearest-neighbor 
interactions. this enables one to maintain an approximate connection 
with the $O(3)$ $\sigma$ model for antiferromagnets in this formalism.
\pr
We now show that
this is helpful for identifying the conformal field theory 
that corresponds to the limiting cases $s(s+1) \rightarrow 0^-$.
In the general case with complex spin,
we start with the 
$\sigma$ model continued analytically to negative $s$:
\be
   \frac{1}{s(s + 1)} \int d^2 z \sum_{i=1}^{3} (\partial _\mu \eta ^i)
g_{ij} (\partial _\mu \eta ^j)
\label{complspin}
\ee
where the 
spin variables satisfy
\be
 \sum _{i=1}^{3} \eta ^i g_{ij} \eta _i  = s(s + 1)
\label{sumspin}
\ee
and the metric $g_{ij} $ that contracts the spin indices is Minkowskian 
in the non-compact case, and Euclidean in the compact case.
The Casimir factor $s(s + 1)$ 
should be retained as one takes the singular
limit $s(s + 1) \rightarrow 0$,
corresponding to the limits $ s  \rightarrow 0^-$ and $ s=-1^+$,
where it becomes a singular constraint that should be solved
without making a prior normalization with respect to $s ( s + 1)$.
Defining the variables
\bea
  w &=& \frac{\eta _1 + i \eta _2 }{a - i\eta _3} \nn \\
 {\overline  w} &=& -\frac{\eta _1 - i \eta _2 }{a + i\eta _3}, 
\qquad a \equiv \sqrt{s(s+1)} \nn \\
\label{changevar}
\eea
where $a^2$ is {\it negative} in the non-compact case, 
at finite $s$ we can map
the classical lagrangian (\ref{complspin})
onto the following $\sigma$ model
\be
    \frac{\lambda^2}{\pi} \int d^2 z \frac{1}{(1 + w {\overline w})^2}(\partial _\mu w
\partial ^\mu {\overline w} )
\label{noncomp}
\ee
where 
\be
    \lambda^2 =\frac{\pi}{s(s+1)}
\label{lambda}
\ee
This resembles formally
a conventional $O(3)$ $\sigma$ model, 
although it has negative-definite metric
in the non-compact case $s(s + 1) <0$.
Formally, the central charge would be 
$c=3k/(k+2)$ with $k=2s$. 
\pr
We observe that
the metric $g (w)$ becomes singular
in the limit $s ( s +1 ) \rightarrow 0$.
and the theory is topological,
although the metric is not
singular for other values of ${s (s + 1)}$. However,
it should also be noted that the expression of the action in terms of the
$\eta$ variables is not regular at the point $\eta _1 =\eta _2 =0$.
To avoid this problem, as we shall discuss below, we
define the theory in the $s ( s +1) <0 $ regime
through analytic continuation, using the variables (\ref{changevar}),
in terms of which we construct
a $\sigma$ model with Minkowski metric
$g(w)=1/(1 + w {\overline w})$. 
When expressed in terms of the $\eta$ variables,
the action assumes the linear form
(\ref{complspin}), up to an overall normalization. However,
the metric is singular, and the theory 
at the core of the singularity is
topological for any $s$ such that $s ( s +1 ) < 0$.
The limit  $s (s + 1) \rightarrow 0^-$ may be
taken smoothly, with the theory
remaining Minkowskian. 
This limit corresponds precisely to the singularity 
of the black-hole metric, and the 
corresponding 
theory is topological.  
In that limit, the theory can be rotated 
without problems to a Euclidean theory that possesses
instantons, as we discuss below.
This is consistent with the above-mentioned equivalence of
the two cases (\ref{case1}),(\ref{case2}) in the limit $s ( s + 1)
\rightarrow 0^{\pm}$.
\pr
To gain formal insight into the nature of the 
relevant conformal field theory in this limit, 
we notice that when $s(s+1)=0$, where  
the $O(3)$ $\sigma$ model (\ref{noncomp}) 
has a singular metric tensor 
$g(w,{\overline w})=1/(0)^2 \rightarrow \infty$. 
one may regulate the theory in this limiting case by defining variables:
\be
  w=\frac{\eta_1 + i \eta_2}{-i\eta_3}, 
{\overline w} =\frac{\eta_1 - i\eta_2}{i\eta_3}
\label{etas}
\ee
and keeping $s(s+1)$ arbitrarily small but non-zero in  
(\ref{sumspin}). 
Then one reproduces (\ref{complspin}) in the limit $s(s+1) \rightarrow 0$ 
using a metric $g(w,{\overline w})$ in (\ref{noncomp}) 
of the form: 
\be 
   g(w, {\overline w}) = \frac{1}{1 + w {\overline w}}
\label{gww} 
\ee
which is of the cigar-like Euclidean
black-hole type discussed in~\cite{Witten}. 
The latter is known to be described by 
a non-compact $SL(2,R)/U(1)$ conformal field theory
and, as mentioned previously, 
the limiting case $s+1 =0$ corresponds to the singularity of this black 
hole. The latter is known~\cite{Witten,eguchi,emntop} 
to be described by a topological world-sheet conformal field theory, 
obtained from the $N=2$ supersymmetric world-sheet $\sigma$-model 
by a suitable twisting which ensures that $c=0$, as mentioned earlier.
\pr
All the above coset constructions require
$U(1)$ symmetries.
As we
have shown in the previous Section, the holomorphic sector of
Lipatov's spin-chain model, which has ${s = -1}$,
has just such a bosonic symmetry in the limit of a large number
of Reggeized gluons, supporting its identification with the
${k \rightarrow 2}$ limit of the black hole $SU(1,1)/U(1)$
model.
\pr
Thus, we reach the remarkable conclusion that
spin systems with $s ( s + 1) \rightarrow 0^-$, corresponding to
complex spin, can be reformulated as topological $\sigma$ models. 
In point of fact, as we argue below, the topological symmetry
is broken by instanton effects that induce a
non-perturbative renormalization-group flow.

\section{Renormalization-Group Flow, Instantons and High-Energy
Scattering}
\pr
We explore in this Section the extent to which the understanding
obtained above of the topological field-theoretical
continuum limit of Lipatov's spin-chain system may cast light
on the nature of high-energy scattering and provide, in particular,
information on the dependence of the
Reggeon intercept on the number of Reggeized gluons.
Our main
tools in this analysis are the renormalization group and
Zamolodchikov's $C$ theorem~\cite{zam}. 
We recall first that renormalization-group
evolution entails a thinning out of the physical
degrees of freedom, which corresponds to a decrease in the
central charge $c$ for unitary models. As we discuss later,
this theorem requires modification for anti-unitary models
such as those relevant to high-energy scattering.
\pr
We start by discussing the unitary models in the ${s \ge 1/2}$ region
of the Figure, which are described by $SU(2)/U(1)$ non-linear
$\sigma$ models, as discussed in Section 3. The effective
coupling $g(L)$ increases as the infrared cutoff $L$ is
increased:
\be
\frac{dg}{dln L} = \frac{g^2}{2\pi} \qquad ; \qquad
g (L) \simeq g_0/[ 1 - \frac{g_0}{2\pi}ln L ]
\ee
Bearing in mind the relation ${g = 2/s}$, we see that this
corresponds to a decrease in the effective spin $s$, i.e., a
decrease in the level parameter ${k = 2 s}$ and hence in the
central charge (\ref{twelve}), in agreement
with Zamolodchikov's $C$ theorem.
\pr
A similar analysis applies in the other unitary region, ${s < -1}$
corresponding to ${k > 2}$ for the $SU(1,1)/U(1)$ non-linear
$\sigma$ model. This region has been discussed elsewhere~\cite{emntop} in
connection with string black-hole decay, which is due to
higher-genus effects that renormalize the effective action.
They provide an absorptive part that is a signature of instability,
increase $k$ and hence decrease the black-hole mass, which
is proportional to ${1/\sqrt(k - 2)}$. This also corresponds to a
decrease in the central charge $c$, as given by equation
(\ref{eighteen}), in
agreement with the $C$ theorem. This higher-genus decay effect
can be represented by instantons in the effective lowest-genus
theory, since these describe transitions between string black holes
of different masses, i.e., different values of $k$ and $c$~\cite{emntop}.
\pr
The discussion of theories with ${ 1/2 > s > -1}$ is more
complicated, because they are anti-unitary, a property
traceable to the fact that in
high-energy scattering one is calculating the energy
dependence
\be
s^{\epsilon} = e^{\epsilon log s} \quad : \quad <H> = \epsilon
\label{logs}
\ee
rather than a normal unitary evolution ${e^{i H t}}$. The latter
is related to the former (\ref{logs}) by ${H \rightarrow i H}$, which
corresponds to a change in sign in ${c = <T T>}$.
Under these circumstances, Zamolodchikov's
$C$ theorem does not apply~\cite{zam}. 
However, even in anti-unitary theories the
renormalization-group flow must be such as to thin out the
physical degrees of freedom~\cite{thin}.
\pr
Symmetry breaking usually arises because
the unbroken phase of the theory has more degrees of freedom
than the broken phase, as is, for instance, the case for
the topological phase of the $N=2$ $\sigma$ models
corresponding to a Wess-Zumino theory
on $SL(2,R)/U(1)$.
In such models, the topological phase consists of an infinity
of non-propagating topological modes of the string. The latter
couple to the propagating string modes as a result of
non-perturbative conformal invariance~\cite{emntop}.
This theory has instantons (holomorphic maps) whose
suppression is not bounded away from zero~\footnote{The metric
of the $SL(2,R)/U(1)$ Euclidean black-hole target space is
actually the
limiting case in which the instantons are unsuppressed, as a result
of the non-compact moduli space.}. These
induce extra logarithmic scale dependences in correlation
functions, vacuum energies, etc.,
which depend on the size
of the world sheet. They imply a breaking
of the topological symmetry and a thinning of the physical
degres of freedom of the system.
\pr
We now argue that a similar instanton effect occurs
in the case of high-energy scattering, starting
from the conventional
representation of the $s > 0$ spin system
in the continuum limit as an $O(3)$ $\sigma$ model.
This representation holds exactly only in the limit of
large $s$, but it will be sufficient
for our purposes. Denoting by
${\eta} _i: i=1,2,3$ the mean-spin variable,
with $|{\eta}| =1 $, the action of the $O(3)$
$\sigma$ model can be written in terms of the complex
variables (\ref{changevar}).
The action $\int d^2 x (\partial _\mu \eta _i )^2 $
can then be written in the form (\ref{noncomp}), 
where 
the metric  $g (w)$ is given by
\be
g (w, {\overline w})=\frac{1}{ (1 + |w|^2)^2}
\label{metricw}
\ee
The $\sigma$ model (\ref{noncomp}) with the metric
(\ref{metricw}) has instanton  solutions
\be
     w(z)= u + \frac{\rho}{z - z_0}
\label{inst}
\ee
with winding number $n=1$~\footnote{It is straightforward
to incorporate solutions with higher winding number,
that have a proportionality constant 
$n$ in front of the instanton action.},
which describe transitions between the
different topological sectors of the theory, that are
classified by the $\theta$ term of the model.
These instanton transitions reduce the central charge $c$
by reducing $k = 2s$ towards zero from above,
as illustrated in the Figure.
\pr
We now extend this discussion
to include non-compact target spaces.
To this end, we generalize the metric $g(w)$
(\ref{metricw}) to
\be
g(w,{\overline w}) = \frac{1}{(1 + |w|^2)^q}
\label{metricq}
\ee
with $q$ arbitrary but real. Instanton solutions
of the classical action exist only for $q > 1/2$.
For $q > 1$ the target space of the $\sigma$
model is compact, and one has the conventional instanton. 
We have argued in the previous section (\ref{gww}) that the 
case $q=1$ corresponds to the
conformal field theory describing the 
limit $s(s + 1) \rightarrow 0$. 
The instanton action is finite~\cite{yung} for $q=1$:
\be
S_I=a \frac{1}{3} \lambda^2, 
\label{instact}
\ee
where $a$ is a numerical coefficient that 
depends on the regularization scheme.
This implies that the instanton contribution in the correlation 
functions of the model will be weighted by $g_I$, where 
\be
      g_I = e^{-a \frac{\lambda^2}{3}}
\label{ind}
\ee
In compact $\sigma$ models, the anti-instantons 
have zero action, and as such they do not contribute to 
correlation functions. On the other hand, in the black-hole 
conformal field theory, the anti-instantons make non-trivial contributions 
to the correlation functions of the model~\cite{yung}.  
Their effects may be summarized by adding an effective vertex 
\be
    V_{{\bar I}} \sim -g_{{\bar I}} \int d^2 \sigma g (w) {\overline \chi}
\chi \partial ^2 _\mu ( g(w,{\bar w}) {\overline \chi}\chi )
\label{antinst}
\ee
where $\chi$ denotes the spin-$0$ fermions of the twisted
$N=2$ supersymmetric black-hole $\sigma$ model.
\pr
The observables 
in the topological $q=1$ $\sigma$ model 
do not depend directly on $\lambda ^2$. In view of the
above-mentioned 
regularization-scheme dependence, therefore, one may consider 
$g_I$ (\ref{ind}) as the true renormalized 
coupling constant of the model~\cite{yung}. 
As already mentioned, the value $q=1$~\cite{yung}
marks the border line between the compact and
non-compact target-space cases, where the moduli space of the instantons
is non-compact.
In the topological version of the
$SL(2,R)/U(1)$ model, the instanton action is finite
and the instantons constitute relevant operators,
as far as the breaking of conformal invariance is
concerned~\cite{emndollar,yung}.
Notice that 
in the limit $s(s+1) \rightarrow 0^+$, which for negative 
$s$ occurs for $s < -1$, the positive instanton 
action (\ref{instact}) becomes infinite, and the coupling constant
$g_I \rightarrow 0$ (\ref{ind}). On the other hand,  
in the region where $s(s+1) < 0$,
the instanton contributions to the correlation functions 
are not suppressed, and in fact diverge as $s(s +1 ) \rightarrow 0^{-}$. 
In this domain of the renormalization-group flow, the instanton 
transitions therefore occur very rapidly. 
\pr
The presence of instanton transitions leads, as we show
below, to a breaking of the topological symmetry~\cite{yung}
in the sense of a false vacuum~\footnote{The presence of a non-zero Witten
index in the twisted $N=2$ supersymmetric $\sigma$ model
implies that supersymmetry can only be broken in the sense of a
false vacuum.}. The presence 
of the false vacuum implies that the phase $s(s+1) < 0$ 
is unstable, driving the theory to the limiting case $s = 0$,
which is 
equivalent to our `principal value' version of the case $s = -1$.
\pr
To this end, we first review the
breaking of the topological symmetry by instantons
in this class of theory. First, we note that in our case
the existence of instantons is associated  with
the Berry-phase term
in the spin model, as discussed in~\cite{shankar}.
For our purpose, we note that this term 
becomes, in the
case of $s=0$, just the complex-structure
term of the topological $\sigma$ model,
i.e., in terms of the $w$ variables (\ref{changevar}),
\be
S_B = \int d^2 z g(w) ({\overline \partial} w \partial
{\overline w} - {\overline \partial} {\overline w}
\partial w )
\label{compl}
\ee
with the same normalization coefficient as the kinetic term.
In terms of the $\eta _i$
spin variables, this yields a term
\be
 \frac{1}{s(s + 1)}
    \int d^2 z \epsilon _{\alpha\beta} \frac{\eta _1}{\eta _3}
\partial _\alpha \eta _2 \partial _\beta  \eta _3
\label{berry}
\ee
which, using the Casimir constraint to express $\eta _3 $
in terms of $\eta_{1,2}$, becomes a total derivative
\be
         \frac{1}{s(s+ 1)}\int d^2 \epsilon _{\alpha\beta}
\frac{1}{\eta _1^2 +  \eta _2 ^2}
\partial _\alpha (\eta _2^3)\partial _\beta \eta _1
\label{parti}
\ee
We note that the Berry-phase term (\ref{berry})
differs from the conventional Berry-phase spin term
by the factor $1/{\eta _3}$. This extra power of the spin variable
is essential in this singular limit to guarantee the correct
dimensionality in spin space. 
To understand this, note that,
in the non-singular case, normalization
of the spin variable by division by the square root of the
non-vanishing Casimir coefficient is possible. However, this is not
possible
in the singular limit we are considering here. In this case, the point
$\eta _i =0$ for all $i=1,2,3$ is allowed, in contrast to the non-singular
positive-spin case. The Berry-phase term has to be regular
at this point, since it is a finite topological
invariant, the winding number,
and the only way to achieve this is to
normalize by dividing with $1/{\eta _3}{s(s + 1)}$.
The guiding principle is to construct a continuous
version of the Berry-phase  term that renders the instanton
deformations of the $\sigma$ model relevant operators.
At this stage, we still lack a first-principles construction
of the complex-structure term from the underlying statistical
model, but the above heuristic
arguments for its form are sufficient for our purposes.
\pr
The existence of such a topological
term guarantees that
the instantons have finite action, in contrast to the
anti-instantons whose action diverges logarithmically 
with the area
of the world-sheet.
Thus instanton-anti-instanton configurations
can lead to extra logarithmic dependences in correlation
functions, that can affect the conformal invariance.
The doping Ansatz we adopted earlier can supersymmetrize the
$\sigma$ model, as appropriate for its topological nature in the limit
$s=0$. For an analysis of instantons in this supersymmetric
version see~\cite{yung}. The important point is that the instantons
result~\cite{emn}
in a renormalization of the Wess-Zumino level parameter
of the $\sigma$-model $k(=2s) \rightarrow k({\rm ln}(\Lambda/\ell )$,
where $\Lambda$ ($\ell $) is an infrared (ultraviolet) 
world-sheet renormalization-group length scale. 

To understand this, we first note that 
the instanton-anti-instanton vertices 
introduce new terms into the effective
action. Making a derivative expansion
of the instanton vertex and taking the large-$k$
limit, i.e., restricting our attention
to instanton sizes $\rho \simeq \ell$, these new terms
acquire the same form as the kinetic terms in 
the $\sigma$ model,
thereby corresponding to a renormalization of
the effective level parameter in the
large $k (=2s)$ limit~\cite{emn}, related to 
the $SL(2,R)/U(1)$ coset black-hole model~\cite{Witten}:
\be
 k(=2s) \rightarrow k - 2\pi k^2 d'
\qquad : \qquad
 d' \equiv g^Ig^{{\overline I}}\int \frac{d|\rho|}{|\rho|^3}
\frac{\ell ^{2}}{[(\rho/\ell)^2 + 1]^{\frac{k}{2}}}
\label{twelve2}
\ee
where $\rho$ denotes the collective coordinate of the instantons
(\ref{inst}). 
If other perturbations are ignored,
the instantons are irrelevant deformations
and conformal invariance is maintained.
However, in the 
$SL(2,R)/U(1)$ coset black-hole model, 
there exist matter deformations,
$T_0 \int d^2z {\cal F}_{-\frac{1}{2}, 0,0}^{c,c}$,
with $T_0$ assumed positive 
in the $SL(2,R)$ notation of \cite{chaudh},
which change drastically the situation~\cite{emn}. 
Similar matter excitations also appear in the spectrum of the 
exact solutions of the 
Baxter equation for the spin model of high-energy QCD   
of \cite{fadeev}, so we need to take them 
into account. 

The matter deformations induce 
extra logarithmic infinities
in the shift (\ref{twelve2}), that are visible in the dilute-gas 
and weak-matter approximations.
In this case, there
is a contribution to the $\sigma$-model effective action of the form
\be
S_{eff} \ni -T_0\int d^2z d^2z'<{\cal F}_{-\frac{1}{2}, 0, 0}^{c,c}
(z,{\bar z}) V_{I{\overline I}} (z',{\bar z}')>
\label{tachdeform}
\ee
where $V_{I{\overline I}}$ denotes the instanton-anti-instanton deformation.
Using the
explicit form of the matter
vertex ${\cal F}$
\bea
{\cal F}_{-\frac{1}{2},0,0}^{c,c}=
\frac{1}{\sqrt{1 + |w|^2}}\frac{1}
{\Gamma (\frac{1}{2})^2}
\sum_{n=0}^{\infty} \{ 2\psi (n+1)
-2\psi (n+\frac{1}{2}) + \nn \\
+ ln(1 + |w|^2) \} (\sqrt{1 + |w|^2})^{-n}
\label{expression}
\eea
given by $SL(2,R)$ symmetry
\cite{chaudh}, it is straightforward
to isolate a logarithmically-infinite contribution
to the kinetic term in the $\sigma$ model, associated
with infrared infinities on the world sheet. 
These are expressible in terms 
of the world-sheet volume $V^{(2)}/\ell ^2=\Lambda ^2 /\ell ^2$,
the latter 
measured in units of the ultraviolet cut-off $\ell $: 
\bea
 S_{eff} &\ni& -T_0g^Ig^{{\overline I}} 
\int d^2z' \int
\frac{d\rho}{\rho} (\frac{\ell ^2}{\ell ^2 + \rho^2})^{\frac{k}{2}}
\int d^2 z \frac{1}{|z-z'|^2}
\frac{1}{1 + |w|^2}
\partial _{z'} w(z')
\partial _{\bar z'} {\overline w}(z') + \dots \nn \\
&\propto &  -T_0g^Ig^{{\overline I}}  {\rm ln} \frac{\Lambda ^2}{\ell ^2} \int d^2z'
\frac{1}{1 + |w|^2}
\partial _{z'} w(z')
\partial _{\bar z'} {\overline w}(z')
\label{analyticexp}
\eea
The logarithmic scale dependence 
can be absorbed in a
shift of $k$: $k_{ren} = k -T_0g^Ig^{{\overline I}}  {\rm ln}\left(\Lambda /\ell \right)$.
The net result of such a renormalization is to reduce the magnitude 
of $k$~\cite{emn}.
The central charge $c=3k/(k-2)$ of the twisted model changes as follows: 
\be
\frac{\partial}{\partial t}c=-\frac{6}{(k-2)^2}\frac{\partial}{\partial t}k
\quad ; \quad t \equiv {\rm ln}\left(\Lambda /\ell \right) 
\label{cfunction}
\ee
Thus, the rate of change of $c$ is {\it opposite } to that of $k$. 
Therefore, by reducing $k$, one increases the central charge. 
Since, for 
$s \rightarrow 0^-$ $(k \rightarrow 0^+)$, $c \rightarrow 0$, one 
then observes 
from (\ref{cfunction}) 
that, 
under the instanton-induced renormalization-group flow,
the central charge
is driven towards $c=0^-$, in the limit $t \rightarrow \infty$,
as $c \simeq -6/(T_0g^Ig^{{\overline I}}t)$. 
\pr
We note now that the resulting vacuum energy
can be found by computing the vertex operator
of an anti-instanton in the dilute-gas approximation for a
$\sigma$-model deformation corresponding to an
instanton vertex operator. The result
to leading order in the instanton-anti-instanton coupling is
\be
E_{vac}^{I {\overline I}}   =   <V_{{\overline I}}>_I=
-g^{{\overline I}}\int d^2 x_1 \partial^2 _{x_2} < O(x_1) O(x_2) >|_{x_1 \rightarrow
x_2}
\label{itwo}
\ee
Recalling that the dominant anti-instanton
configurations have sizes ${\rho \simeq \ell }$, we can use
(\ref{itwo}) to estimate that in the infrared limit
when ${\Lambda/\ell  >> u}$
\be
E_{vac} =-16\pi^2 g^I g^{{\overline I}}
\frac{V^{(2)}}{\ell ^2}[ ln(\Lambda /\ell ) + O(1)]
\label{ithree}
\ee
where $V^{(2)}$ is the world-sheet volume.
\pr
At this stage, we appear to have some logarithmic
dependence as a result of instanton
configurations.
However, as was argued in~\cite{yung},
upon summing over an arbitrary number of 
instanton-anti-instanton pairs in the model, 
the logarithmic infrared divergences in (\ref{ithree})
disappear, and the system is equivalent to a Coulomb-gas/sine-Gordon 
model, in a similar spirit to the compact $O(3)$ case,  
the formal difference from the latter being that the r\^ole of instantons
in that case is now played by the instanton-anti-instanton pairs.  
\pr
The details of the analysis can be found in~\cite{yung}, and we
describe here only the basic results that are relevant for our purposes. 
When one maps the system, resummed over an arbitrary number of instantons
and anti-instantons, to a Coulomb gas, the resulting vacuum energy 
(\ref{ithree}) may be re-calculated using 
the free massive-fermion representation~\cite{yung}, with action:
\be
   S_{eff} =\frac{1}{\pi} \int d^2x 
\{ {\overline \psi} \gamma _\mu \partial_\mu \psi 
+ m {\overline \psi} \psi \}
\label{free}
\ee
where the 
mass $|m|=\frac{2\pi}{\ell }\sqrt{8\pi}\sqrt{g_I g_{{\overline I}}}$,
whose inverse plays 
the r\^ole of a world-sheet infrared cut-off for the system. 
The resulting vacuum energy
is now given by: 
\be
    E_{vac} =\frac{m^2}{2\pi} V^{(2)} {\rm log}\frac{m\ell }{2}
\propto V^{(2)} g_I g_{{\overline I}}
\left({\rm log}(g_I g_{{\overline I}}) + {\rm const} \right)
\label{irfree}
\ee
which shows that the 
model has finite vacuum energy
when resummed over the instanton-anti-instanton pairs,
which still break the topological symmetry~\cite{yung}. 
We note that 
the vacuum energy becomes zero 
in the limit where $s(s+1) \rightarrow 0$ such that $s(s+1)>0$, and the
topological symmetry at the singularity 
is restored. 
\pr
In the limit of a large number of Reggeized gluons, the spatial size
of the system, corresponding to the volume of the
impact-parameter space, can be related to the 
number of Reggeons~\footnote{We are back to the half-filled
case, where the charge (hole) excitations
acquire a topological nature. In the $s=0$ case, this is
also true for the spin excitations as well.}: $V^{(2)} \propto N_g$.
Thus the vacuum energy of our problem
exhibits a non-trivial dependence on the
number of Reggeized gluons $N_g$ in the scattering
process, given in the large-$N_g$ limit by
\be
 E_{vac} \propto N_g 
\label{intercept}
\ee
where we have used fact that the world-sheet
volume is proportional to $N_g$.
Taking into account also the relation~\cite{Lipatov} between $E_{vac}$
and the Regge intercept $j $:
 $E_{vac} = 1 - j = -\Delta  $, we see that the Regge
intercept varies linearly
with increasing $N_g$, at least for large $N_g$.
A similar result has also been argued on the basis of a Hartree-Fock
approximation to Lipatov's Hamiltonian~\cite{braun}. 
The sign of the vacuum energy and hence the shift in the
Regge intercept is currently ambiguous 
in our approach, because 
the instanton coupling constant (\ref{ind})
has a regularization-scheme dependence~\cite{yung}, and hence
should be considered as arbitrary within our approach.
For positive coefficients $a$ and $s(s+1) 
\rightarrow 0^-$, the 
instanton coupling 
$g_I \rightarrow \infty$, and one would obtain positive (infinite) 
vacuum energy (\ref{irfree}). However, there   
exist regularization schemes such that $a$ is proportional 
to $s(s+1)$ in such a way so that $g_I$ is finite, and even smaller than one. 
In the framework of high-energy QCD, 
such ambiguities 
may be associated with the renormalization-group running
of the coupling constant of the system of $N_g$ gluons,
which according 
to \cite{braun} could be responsible for the appearance of 
negative ground state energies 
in the 
Hartree-Fock approximation. In contrast, in the fixed coupling constant 
approach to multicolour QCD, on which the integrable model analysis of 
\cite{fadeev} is based, the energy comes out positive, implying the 
instability of multicolour states, which thus become irrelevant 
at high energies.
\pr
We conclude that the instantons 
induce an instability
and break the topological symmetry~\cite{yung} via 
a false vacuum. 
This results in a tendency of the $-1 < s < 0$  
system to flow towards the $s=0$ ground case. 
The appropriate choice of vacuum state for the
ambiguous case $s=-1$ is then fixed by the requirement of
isomorphism to the holomorphic sector,
as argued in \cite{fadeev}.
\pr
\section{Conclusions and Prospects}
\pr
We have analyzed in this paper the symmetries of 
Lipatov's model for high-energy scattering in QCD,
and used them to motivate a proposal for the
conformal field theory that should describe the continuum limit of
Lipatov's model corresponding to the exchange of a large number of
Reggeized gluons in the $t$ channel. Arguing by analogy with the known
correspondence between compact Heisenberg spin-chain models and
non-linear $O(3)$ $\sigma$ models, we have suggested that Lipatov's
model may corresond to a limiting case of a non-compact
$SL(2,R)/U(1)$ $\sigma$ model. An analysis of instantons helps to explain
the appearance of this
limiting model, which is a topological
field theory analogous to that describing the core of a 
$1+1$-dimensional string black hole. It possesses an $N=2$
supersymmetric algebra that includes the $W_{\infty} \otimes W_{\infty}$
bosonic algebra previously identified in Lipatov's model.
Formal support of our proposal for identifying the 
conformal field theory underlying Lipatov's model as the $SL(2,R)/U(1)$ 
model is provided by the observation reported in the second paper
of~\cite{fadeev}, that
the exact solution of the Baxter equation 
for the $N_g=2$ Reggeon state bears a remarkable similarity 
to the spectrum of the $SL(2,R)/U(1)$ coset conformal field theory. 
Our spin-charge-separation Ansatz may extend this similarity 
to the multi-Reggeon case as well. 
\pr
Many aspects of our analysis are heuristic, and merit
further study. These include the validity of the `hole-regulator'
scheme that we have proposed, the quantum corrections to the
$W_{\infty} \otimes W_{\infty}$ symmetry algebra that we have
identified, details of its elevation from `world sheet' to
`target space', and the representation of the second
$W_{\infty}$ algebra. The relation of the non-compact
spin-chain and $\sigma$ models should be clarified, as has
previously been done for the compact spin-chain models and
$O(3)$ $\sigma$ models. Also, the r\^ole of instantons in
non-compact $\sigma$ models merits further investigation.

\pr
We hope that our proposal may open the way to a more
powerful tool-box for analyzing high-energy scattering in QCD.
Field-theoretical techniques may allow the consequences of
both $t$- and $s$-channel unitarity to be investigated
more thoroughly, via the string topological diagram
expansion and the power of conformal field theory.
\pr
\nk {\Large{\bf  Acknowledgements}}
\pr
We thank Ioannis Bakas, Merab Eliashvili,
Ludwig Faddeev and Gregory Korchemsky for useful
discussions. 
J.E. thanks the Royal Nepalese Academy of Science
and Technology and Tribhuvan University for their hospitality
while this work was being started, and the Rutherford-Appleton
Laboratory for its hospitality while it was being completed.
Our interest in non-compact non-linear $\sigma$ models arose
from our work with D.V. Nanopoulos on their relevance
to black-hole problems, and we acknowledge with pleasure the
insights gained during that collaboration.
\pr


\begin{thebibliography}{99}
\bibitem{Lipatov} L.N. Lipatov, 
Proceedings of  {\it Perturbative QCD}, ed. A.H. Mueller (World Sci.,
Singapore, 1989), 411; J.E.T.P. Lett. 59 (1994), 596.
\bibitem{fadeev} L.D. Faddeev and 
G.P. Korchemsky,
Phys. Lett. B342 (1995), 311; \\
G.P. Korchemsky, Nucl. Phys. B443 (1995), 255. 
\bibitem{odderon} G.P. Korchemsky, hep-ph/9801377;
\par R.A. Janik and J. Wosiek, hep-th/9802100;
\par M. Praszalowicz and A. Rostworowski, hep-ph/9805245;
\par M.A. Braun, hep-ph/9805394. 
\bibitem{braun} M. Braun, Phys. Lett. B348 (1995), 190;
{\it ibid.} B351 (1995), 528. 
\bibitem{2to4} J. Bartels, L.N. Lipatov and M. W\"usthoff, 
Nucl. Phys. B464 (1996), 298.
\bibitem{affleck} I. Affleck, Lectures given at the
{\it Summer School on Fields, Strings and Critical Phenomena},
Les Houches, France, Jun 28 - Aug 5, 1988; and references therein.  
\bibitem{Hall} M.S. Girvin and R.A. Prange, {\it The Quantum 
Hall Effect} (Springer, New~York, 1990). 
\bibitem{Witten} E. Witten, Phys. Rev. D44 (1991), 314.
\bibitem{shankar} R. Shankar, Nucl. Phys. B330 (1990), 433.
\bibitem{dorey} N. Dorey and N.E. Mavromatos,
Phys. Lett. B250 (1990), 107; Phys. Rev. B44 (1991), 5286; \\
for a comprehensive review, see: N.E. Mavromatos, Nucl. Phys. 
Proc. Suppl. C33 (1992), 145.
\bibitem{bakir} I. Bakas and E. Kiritsis, Prog. Theor. Phys. Suppl. 102
(1990), 15; Int. J. Mod. Phys.  A7 (1992) 55;
and references therein. 
\bibitem{susy} P.B. Wiegmann, { Phys. Rev. Lett.} { 60} 
(1988), 821; \\
S. Sarkar, {J. Phys.} { A23} (1990), L409; 
{ J. Phys. } {A24} (1991), 1137; \\
F.H.L. Essler, V.A. Korepin and K. Schoutens, { Phys. Rev. Lett.}
{68} (1992), 2960; \\
A. Lerda and S. Sciuto, {Nucl. Phys. } {B410} (1993), 577.  
\bibitem{WW} E. Bergshoeff, C.N. Pope, L.J. Romans, E. Sezgin and 
X.G. Shen, Phys. Let. B245 (1990), 447. 
\bibitem{mandal} S. R. Wadia, hep-th/9411213; \\
A. Dhar, G. Mandal and S. R. Wadia, 
Phys. Lett. B329 (1994), 15.  
\bibitem{emntop} J. Ellis, N.E. Mavromatos and D.V. Nanopoulos, 
Phys. Lett. B228 (1992), 23. 
\bibitem{eguchi} T. Eguchi, Mod. Phys. Lett. A7 (1992), 85. 
\bibitem{lykken} L. J. Dixon, M. E. Peskin and  J. Lykken, 
Nucl.Phys.B325 (1989), 329.  
\bibitem{emndollar} J. Ellis, N.E. Mavromatos and D.V. Nanopoulos, 
Mod. Phys. Lett. A10 (1995) 425; hep-th/9305117. 
\bibitem{yung} A.V. Yung, Int. J. Mod. Phys. A10 (1995), 1553;
{\it ibid.} A11 (1996), 951. 
\bibitem{zam} A.B. Zamolodchikov, J.E.T.P. Lett. 43 (1986), 731;
Sov. J. Nucl. Phys. 46 (1987), 1090;
\par A.A.W. Ludwig and J.L. Cardy, Nucl. Phys. 
B285 [FS19] (1987), 687.
\bibitem{thin} D. Kutasov, Mod. Phys. Lett. A7 (1992), 2943.
\bibitem{emn} J. Ellis, N.E. Mavromatos and D.V. Nanopoulos, 
Lectures presented at the Erice 
Summer School, 31st Course: {\it From 
Supersymmetry to the Origin of Space Time}, Ettore Majorana Centre,
Erice, July 4-12 1993; 
hep-th/9403133 and {\it Proc. Subnuclear Series} 
(World Scientific, Singapore 1944), Vol. 31, p. 1. 
\bibitem{chaudh} S. Chaudhuri and J. Lykken, Nucl. Phys B396 (1993),
270.

\end{thebibliography}
\end{document}